\documentclass[times, 10pt,twocolumn]{article} 
\usepackage{latex8}
\usepackage{times}
\usepackage{graphics}
\usepackage{hhline}
\usepackage{verbatim}

\begin{document}

\title{Frequent itemsets mining for database auto-administration}

\author{
Kamel Aouiche, J\'{e}r\^{o}me Darmont\\
ERIC/BDD, University of Lyon 2\\
5 avenue Pierre Mend\`{e}s-France\\
69676 Bron Cedex, FRANCE\\ \{kaouiche, jdarmont\}@eric.univ-lyon2.fr\\
\and
Le Gruenwald\\
School of Computer Science\\University of Oklahoma\\
Norman, OK 73019, USA\\ ggruenwald@ou.edu\\
}

\maketitle
\thispagestyle{empty}

\begin{abstract}
With the wide development of databases in general and data warehouses in particular,
it is important to reduce the tasks that a database administrator must perform
manually. The aim
of auto-administrative systems is to administrate and adapt themselves
automatically without loss (or even with a gain) in performance. The idea of
using data mining techniques to extract useful knowledge for administration
from the data themselves has existed for some years. However, little
research has been achieved. This idea nevertheless remains a
very promising approach, notably in the field of data warehousing, where
queries are very heterogeneous and cannot be interpreted easily. The aim of
this study is to search for a way of extracting useful knowledge from stored data themselves
 to automatically apply performance optimization techniques,
and more particularly indexing techniques. We have designed a tool that
extracts frequent itemsets from a given workload to compute an index
configuration that helps optimizing data access time. The experiments we
performed showed that the index configurations generated by our tool allowed
performance gains of 15\% to 25\% on a test database and a test data warehouse.
\end{abstract}

\section{Introduction}

Large-scale usage of databases requires a Database Administrator (DBA) whose principal
role is data management, both at the logical level (schema definition) and
the physical level (files and disk storage), as well performance
optimization. With the wide development of
Database Management Systems (DBMSs), minimizing the administration function
has become critical to achieve acceptable response times even at load peaks~\cite{WEI02}. 
One important DBA task is the selection of
suitable physical structures to improve the system performances by
minimizing data access time~\cite{FinkelsteinST88}.

Indexes are physical structures that allow a direct access to the data.  From the
DBA's point of view, performance optimization lies mainly in the selection of indexes and
materialized views~\cite{Agrawal2001,Gupta99}. These physical structures play
a particularly significant role in decision-support databases such as data
warehouses due to their huge volume and
complex queries.

The problem of selecting an optimal index set for a database has been studied since
the seventies. The most recent studies regarding index selection  
use the DBMS' query optimizer to estimate the cost of
various configurations of candidate indexes~\cite{AgrawalCN00,chaudhuri97efficient,ChaudhuriN98,FrankON92}.
However, the idea of using data mining techniques to extract useful knowledge for
administration from the data themselves has been around for some years~\cite{Chaudhuri98}. 
Little work has been done, though. In this paper, we designed and coded a tool that exploits data mining to
recommend a relevant index configuration.

Assuming that index utility is strongly correlated to the usage frequency
of the corresponding attributes within a given workload, the search
for frequent itemsets~\cite{agrawal93mining} appeared well adapted to highlight
this correlation and facilitate index selection. Our tool parses the transaction log file 
(the set of queries executed by the DBMS) to build a context for mining frequent itemsets.
This context connects queries from the input workload to the attributes that may be indexed.
The output frequent itemsets are sets of attributes forming a configuration of candidate
indexes. Finally, various strategies can be applied to select the indexes to 
effectively build from within this configuration. 

In the remainder of this paper, we present our proposal in Section~\ref{proposal} and
some preliminary experimental results in Section~\ref{exp}, and then finally
conclude the paper and present future research perspectives in 
Section~\ref{conclusion}.

\begin{figure*}[bt]
\begin{center}
\begin{tabular}{|p{13.5cm}|}
    \hline
    {\small {\bf Q1:} {\tt SELECT * FROM T1, T2 WHERE A BETWEEN 1 AND 10 AND C=D}}\\
    {\small {\bf Q2:} {\tt SELECT * FROM T1, T2 WHERE B LIKE '\%this\%' AND C=5 AND E$<$100}}\\
    {\small {\bf Q3:} {\tt SELECT * FROM T1, T2 WHERE A=30 AND B$>$3 GROUP BY C HAVING SUM(E)$>$2}}\\
    {\small {\bf Q4:} {\tt SELECT * FROM T1 WHERE B$>$2 AND E IN (3, 2, 5)}}\\
    {\small {\bf Q5:} {\tt SELECT * FROM T1, T2 WHERE A=30 AND B$>$3 GROUP BY C HAVING SUM(E)$>$2}}\\
    {\small {\bf Q6:} {\tt SELECT * FROM T1, T2 WHERE B$>$3 GROUP BY C HAVING SUM(E)$>$2}}  \\
    \hline
\end{tabular}\caption{Sample workload}\label{fig:Workload example}
\end{center}
\end{figure*}

\section{Frequent itemsets mining for index selection}
\label{proposal}

\subsection{Principle}

Our approach exploits the transaction log to extract an index
configuration.  The queries from the transaction log  constitute a
workload that is treated by an SQL query analyzer. The SQL query analyzer
extracts all the attributes that may be indexed (indexable attributes). Then, we
build a ``query-attribute" matrix, the rows of which are the workload queries, and the columns are the
indexable attributes. The role of this matrix
is to link each indexable attribute to the workload queries
it appears in.

This matrix represents the extraction context for frequent itemsets.  To compute these frequent itemsets,
we selected the Close algorithm~\cite{pasquier99discovering,close99is}, because its
output is the set of the frequent closed intemsets (closed regarding the Galois connection~\cite{close99is}),
which is a generator for all the frequent itemsets and their support. In most cases,
the number of  frequent closed itemsets is much lower than the total number of
frequent itemsets obtained by classical algorithms such as Apriori~\cite{agrawal94fast}. 
In our context, using Close enables us to obtain a
smaller (though still significant) configuration of candidate indexes faster.

Finally, we select from the  configuration of candidate indexes (that corresponds
to the input workload) the most relevant indexes and  create them.

\subsection{Workload extraction}

We assume that a workload similar to the one presented in
Figure~\ref{fig:Workload example} is available.
Such a workload can be easily obtained either from the DBMS'
transaction logs, or by running an external application such as Lumigent's
Log Explorer~\cite{logexplorer}.

\subsection{Indexable attributes extraction}

To reduce response time when running a database query, it is best to build indexes on
the very attributes that are used to process the query. These attributes 
belong to the WHERE, ORDER BY, GROUP BY, and HAVING clauses of SQL queries~\cite{chaudhuri97efficient}.

We designed a syntactic analyzer that is able to operate on any SQL query type
(selections and updates --- subqueries are allowed), and extracts all the indexable
attributes. This process is applied to all
the queries from the workload.

\begin{comment}
We designed a syntactic analyzer that is able to operate on any SQL query type
(selections and updates --- subqueries are allowed), and extracts all the indexable
attributes. For example, analyzing the query from Figure~\ref{fig:workload query example}
returns the following attributes: {\em part.partkey, lineitem.partkey,
part.brand, part.container, lineitem.quantity}.  This process is applied to all
the queries from the workload.

\begin{figure*}[t]
\begin{center}
\begin{tabular}{|p{13.5cm}|}
    \hline
{\small {\tt    
    SELECT SUM(lineitem.extendedprice) / 7.0 FROM lineitem, part \newline
    WHERE \underline{part.partkey} = \underline{lineitem.partkey} \newline
    AND \underline{part.brand} = ':1' AND \underline{part.container} = ':2'\newline
    AND \underline{lineitem.quantity} $<$ (\newline
\hspace*{1cm}    SELECT 0.2 * AVG(lineitem.quantity) FROM lineitem \newline
\hspace*{1cm} WHERE \underline{lineitem.partkey} = \underline{part.partkey} )}}\\
    \hline
\end{tabular}\caption{Sample SQL query}\label{fig:workload query example}
\end{center}
\end{figure*}
\end{comment}

\subsection{Building the extraction context for the frequent closed itemsets}

We build a matrix (Figure~\ref{fig:Extraction context}) the rows of which
represent the workload queries, and the columns represent the set of all the indexable
attributes identified in the previous step. This ``query-attribute" matrix
links each query to the indexable attributes within it. 
Attribute presence in a query is symbolized by 1, and absence by 0.

\begin{figure}[hbt]
\begin{center}
\begin{tabular}{l|c|c|c|c|c|} \hhline{~-----}
     & \multicolumn{5}{c|}{\bf Attributes} \\ \hline
    \multicolumn{1}{|l|}{{\bf Queries}} & A & B & C & D & E \\ \hline
    \multicolumn{1}{|l|}{Q1} & 1 & 0 & 1 & 1 & 0 \\ \hline
    \multicolumn{1}{|l|}{Q2} & 0 & 1 & 1 & 0 & 1\\ \hline
    \multicolumn{1}{|l|}{Q3} & 1 & 1 & 1 & 0 & 1\\ \hline
    \multicolumn{1}{|l|}{Q4} & 0 & 1 & 0 & 0 & 1\\ \hline
    \multicolumn{1}{|l|}{Q5} & 1 & 1 & 1 & 0 & 1\\ \hline
    \multicolumn{1}{|l|}{Q6} & 0 & 1 & 1 & 0 & 1\\ \hline
\end{tabular}\caption{Sample extraction context}\label{fig:Extraction context}
\end{center}
\end{figure}

\subsection{Frequent closed itemsets mining}

The Close algorithm scans in breadth first a lattice of closed itemsets in order
to extract the  frequent closed itemsets and their support. Its input is an extraction
context such as the one presented in Figure~\ref{fig:Extraction context}. 

Intuitivelty, a closed itemset is a maximal set of items (attributes) that are common to a set of
transactions (queries). For instance, in Figure~\ref{fig:Extraction context}'s extraction context,
the BCE itemset is closed because it is the largest set of common attributes for the set of queries
\{Q2, Q3, Q5, Q6\}. On the other hand, the BC itemset is not closed since all the queries containing
attributes B and C (Q2, Q3, Q5, and Q6) also contain attribute E.
Eventually, a closed itemset is said frequent when its support is greater or equal to
a threshold
parameter named {\em minsup} (minimal support).

The application of Close on the context presented in Figure~\ref{fig:Extraction context} 
outputs the following set of frequent closed itemsets (and their support) for a
minimal support equal to 2/6: \{(AC, 3/6), (BE, 5/6), (C, 5/6), (ABCE, 2/6),
(BCE, 4/6)\}. We consider this set as our  configuration of candidate indexes.

\subsection{Indexes construction}\label{construction}

The higher the size of the input workload is, the higher the number of candidate 
indexes obtained with our approach becomes. Thus, it is not feasible
to build all the proposed indexes. Index creation time, and later update 
time, would both be too costly. Hence, it is necessary to devise
filtering methods or processes to reduce the number of indexes to 
generate.

The first naive method is to build all the candidate indexes.  This
method is only applicable when the number of indexes is relatively small. In that particular case, 
creation and update times remain acceptable.

In the context of decision-support databases, and more particularly, of data warehouses,
building indexes is a fundamental issue because of the huge
volume of data stored in fact tables and some dimension tables. Thus, it 
is more critical to build indexes on large tables.  Index
contribution on small tables can indeed prove negligible, and even
sometimes, costly.

Statistical input, such as the cardinality of the attributes to be
indexed, may also be exploited to build indexes. An attribute's cardinality
is the distinct number of values for this attribute within a given relation.  
Depending on the cardinality, indexing may be more or less efficient.
If the cardinality is very large, an index degenerates toward a
sequential scan (of the index structure itself); and if it is very small, an index might not bring 
a very significant improvement~\cite{GV99}. Hence, the best choice might be to build indexes on attributes with
an ``average" cardinality.

In this first study, we took a particular interest in table sizes.
We indeed established two strategies to build indexes from the union
of the frequent closed itemsets provided by Close.
The
first strategy systematically builds all the proposed indexes (naive
method).  In this case, each frequent closed itemset corresponds to an index to be created.
The second strategy takes the size of the tables an index refers to into account.  
In this case, the DBA must define whether a table
is large or not, and only indexes on attributes from these large tables are built.

\subsection{Comparison with the existing methods}

Unlike the index selection methods that have been recently developed, the tool that we propose does not communicate
with the DBMS' query optimizer. The communication between the index selection tool and
the optimizer is usually costly and must be minimized.  
An index configuration computing time
is indeed proportional to the size of the workload,
which is typically large.  
Our method based on frequent itemsets mining is also
greedy in terms of computing time, but it is currently difficult for us to determine
which approach generates the heaviest overhead for the system.

However, we are more interested in the quality of the generated indexes.  
For instance, the {\em Index Selection Tool} (IST) developed by Microsoft
within the SQL Server DBMS~\cite{chaudhuri97efficient} exploits a given workload and provides a configuration of
mono-attribute candidate indexes. A greedy algorithm selects the best indexes
from this configuration, using estimated costs computed by the query
optimizer. The process then reiterates to generate two-attribute indexes
using the mono-attribute indexes, and similarly, to generate
multi-attribute indexes of higher order.
By mining frequent closed itemsets, our tool directly extracts a set of  mono-attribute
{\em and} multi-attribute indexes.  Hence, we do not build an initial
mono-attribute index configuration {\em a priori}, and we do not need to use any heuristic to build
multi-attribute candidate indexes by successive iterations like IST.  We
believe that this approach avoids the generation and cost evaluation of irrelevant indexes.

\section{Experiments}
\label{exp}

In order to validate our approach, we have applied it on a test database and a test data
warehouse.  Our objective here is more to find out whether our proposal makes sense in
practice than to perform true performance tests.

We have chosen the TPC-R decision-support benchmark~\cite{tpcr99} for
our experiments on a relational database because it is a standard that should
allow us to easily compare our approach to the other existing
methods in the future. We have generated the TPC-R 1~GB database and
used the benchmark's 22 read-only queries (labeled Q1 to Q22). In this first experiment, we
suppose refresh operations occur off-line. However,
in order to take index management overhead into account, future performance tests 
will also include TPC-R's RF1 and RF2 refresh functions.

On the other hand, to the best of our knowledge, there is no standard benchmark for
data warehouses yet (TPC-DS is still in development~\cite{poess02}).
Hence, we worked on a small datamart that had been previously developed in our laboratory.
This accidentology datamart is composed of an
{\em Accident} fact table and four dimension tables: {\em Place, Condition, Date and
PersonResponsible}.  It occupies 15~MB on disk.
Starting from our previous analyses on this datamart, we also designed a realistic decision-support 
workload that is specifically adapted to it.  
This workload includes both selection and update operations.
We cannot present it in detail here due to lack of space; interested readers are referred 
to~\cite{accidento}.

Both the TPC-R database and the
accidentology datamart have been implanted within the SQL Server 2000 DBMS.

\begin{figure*}[htb]
\begin{center}
\begin{tabular}{|p{13.5cm}|}
    \hline
{\small {\tt 
// Cold run (no timing)\newline
     \underline{FOR} each query in the workload  \underline{DO} \newline
 \hspace*{1cm}     Execute current query \newline
 \underline{END FOR} \newline
// Warm run\newline
\underline{FOR} i = 1 \underline{TO} number\_of\_replications \underline{DO} \newline
\hspace*{1cm}     \underline{FOR} each query in the workload  \underline{DO} \newline
\hspace*{2cm}     Execute current query \newline
\hspace*{2cm}     Compute response time for current query \newline
\hspace*{1cm} \underline{END FOR} \newline
    \underline{END FOR} \newline
    Compute global mean response time and confidence interval}}\\
    \hline
\end{tabular}
\caption{Test protocol}\label{fig:protocol}
\end{center}
\end{figure*}

The test protocol we adopted is presented in Figure~\ref{fig:protocol}.  This
algorithm has been executed for various values of the Close {\em minsup} (minimal support) parameter.
In practice, this parameter helps us
limiting the number of indexes to generate by selecting only those that are
the most frequently used by the workload.  At each step corresponding to a value of  {\em minsup},
we compute the mean response time for the input workload.

\subsection{Experiments on TPC-R}

\begin{figure}[hbt]
{\centering \resizebox*{0.48\textwidth}{!}{\includegraphics{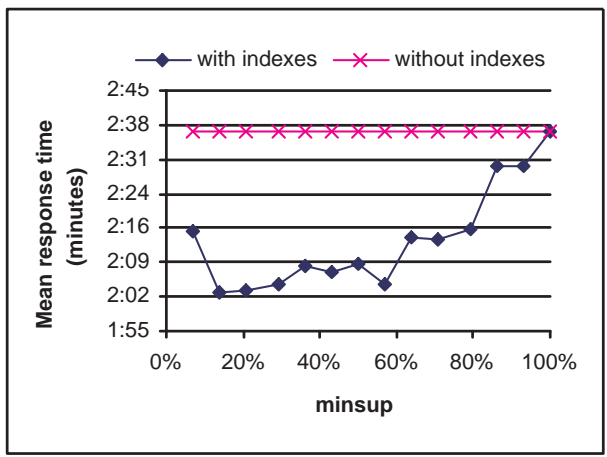}}
\par}
\caption{TPC-R results --- All indexes} \label{fig:TPC-R all index}
\end{figure}

The results we obtained are presented in Figure~\ref{fig:TPC-R all index} 
and~\ref{fig:TPC-R buckley tables}. 
The results from Figure~\ref{fig:TPC-R all index}
correspond to the creation of all the candidate indexes obtained with Close, while the results of
Figure~\ref{fig:TPC-R buckley tables} correspond to a filter on this configuration
(indexes on large tables only; cf. Section~\ref{construction}). 

\begin{figure}[hbt]
{\centering \resizebox*{0.48\textwidth}{!}{\includegraphics{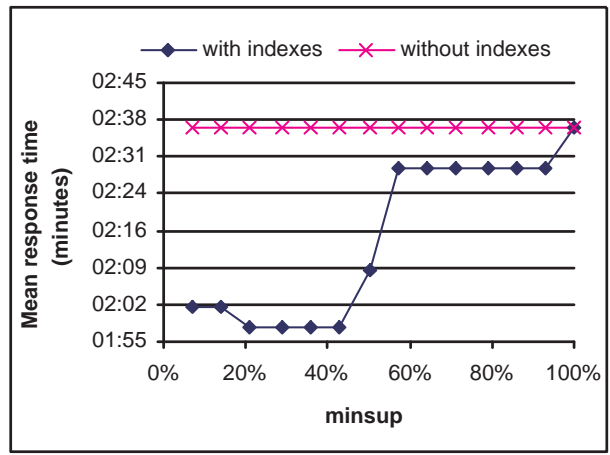}}
\par}
\caption{TPC-R results --- Indexes on large tables} \label{fig:TPC-R buckley
tables}
\end{figure}

Figure~\ref{fig:TPC-R all index} 
and~\ref{fig:TPC-R buckley tables} show that, in comparison with a sequential scan with
no indexes, we achieve a gain in performance for the two strategies regardless of the 
value of {\em minsup}. 
The maximum response time gain, which is
computed as the difference between the mean response time without indexes
and the lowest mean response time with indexes, is close to 22\% in the first
case and 25\% in the second case.  The average response time gains computed
over all values of minsup are 14.4\% and 13.7\% in the first and second case,
respectively.
In the first case, the response time improves until {\em minsup} reaches 15\%, and then degrades at a steady rate,
while in the second case, it remains at its lower value in a broader range (from 20\% {\em minsup} to
50\% {\em minsup}) before degrading abruptly.
The large number
of indexes to be generated in the first case 
can explain this behavior. Considering only
indexes associated with large tables helps reducing the number of generated
indexes and avoids index creation for small tables (since they 
induce a low benefit). 

Finally, for high values of {\em minsup}, the mean response
time becomes close to that obtained without generating any index in both cases. This was
predictable since for a very high {\em minsup}, no or very few indexes are
actually generated. In the second case, this state is reached sooner since fewer indexes are built, which
explains the lower average gain.

\subsection{Data warehouse experiments}

For this series of experiments, we applied the same protocol 
(Figure~\ref{fig:protocol}). However, we did not employ the large table index creation strategy
since all the tables in our test datamart have similar sizes.

\begin{figure}[hbt]
{\centering \resizebox*{0.48\textwidth}{!}{\includegraphics{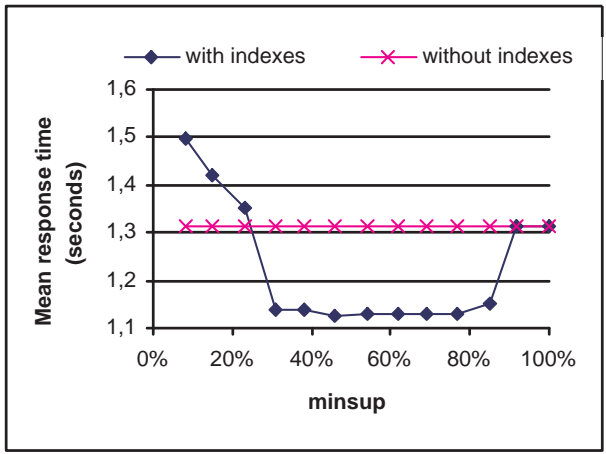}}
\par}
\caption{Accidentology datamart results} \label{fig:Accidentology data mart}
\end{figure}

The results we obtained are presented in Figure~\ref{fig:Accidentology data mart}.
The maximum gain in performance is close to 15\% while the average gain
is 6.4\%. Figure~\ref{fig:Accidentology data mart} 
shows that building indexes is actually more costly than not building them
for {\em minsup}
values ranging between 10\% and 25\%. This may be explained by the high
number of generated indexes, and thus a high index generation time.  
Furthermore, since the 15~MB datamart
is stored completely in
the main memory, the indexes are useful only when it is first loaded.  In this
context, many sparsely used indexes must also be loaded, which penalizes global performance.  

The best gain in response time appears for {\em minsup} values ranging between 30\% and 85\%,
when the number of indexes is such that the index generation overhead is lower than 
the performance increase achieved when loading the datamart.  Beyond  that point,
response time degrades and becomes close to that obtained without indexes because
there are a few or no indexes to generate. 

Another possible explanation
to the lower performances obtained for our datamart, in comparison to the results achieved on
the TPC-R database, may come from the structure of the created indexes. 
Bitmap and star-join indexes are best adapted to data
warehouses~\cite{ONeil95,oneil97improved}, while the default indexes
in SQL Server are variants of B-trees.

\section{Conclusions and perspectives}
\label{conclusion}

We presented in this paper a novel approach for automatic index selection
in DBMSs. The originality of our study rests on the extraction of frequent
itemsets to determine an index configuration. Indeed, we assume that the
importance of an indexable attribute is strongly correlated with its
frequency of appearance in the workload's queries. In addition, the use of a
frequent itemsets mining algorithm such as Close enables us to generate
mono-attribute and multi-attribute indexes on the fly, without having to
implement an iterative process that successively creates increasingly large
multi-attribute indexes based on an initial set of mono-attribute indexes.

Our first experimental results show that our technique indeed allows
response time improvements of 20\% to 25\% for a decision-support workload applied to a relational
database (TPC-R benchmark). We also proposed two strategies to carry
out an index selection among the candidate indexes: the first strategy systematically creates
all the candidate indexes, while the second only creates  the
indexes that are related to so-called large tables. The second strategy allows better
performance improvements because it proposes a better compromise between the space occupied by the
indexes (the number of created indexes is limited to those that are defined on
attributes from large tables) and the use of creating an index (it is
not beneficial to create an index on a small table).

We also performed tests on an accidentology datamart, on which we applied an ad
hoc decision-support workload.  The gain in response time, about 14\%, is less
significant than in the case of TPC-R. This can be explained by the fact that 
the default indexes created by SQL Server are B-tree variants and not bitmap
or star-join indexes, which would be better adapted for a data warehouse.

Our study shows that using data mining techniques for DBMS
auto-administration is promising.  However, it is only a first approach and it
opens up many prospects for research.  Our first research perspective is to improve
index selection by designing more elaborated strategies than the exhaustive
use of a configuration or the exploitation of relatively basic information
relating to table sizes.  A more accurate cost model regarding table features
(other than size), or a strategy for weighting the workload's
queries (by type of query:  selection or update), could help us.  
The use of other unsupervised data mining methods such as
clustering could also provide smaller sets of frequent itemsets.

It also appears essential to test our method further to better
evaluate the overhead it generates, both in terms of
indexes generation time and maintenance time. In particular, it is necessary to apply it on
large data warehouses, while exploiting adapted indexes.  
It would also be very interesting to compare it in a
more systematic way to the IST tool that has been developed by Microsoft, either
through complexity computations of the index configuration generation heuristics (overhead), or
by experiments aiming at evaluating the quality of these configurations (response time
improvement and overhead due to index maintenance).

Extending or coupling our approach with other performance optimization techniques
(materialized views, buffer management, physical clustering, etc.) also constitutes
a promising research perspective.  Indeed, in the context of data warehouses, it is
mainly in conjunction with other physical structures (primarily materialized
views) that indexing allows the most significant performance gains~\cite{AgrawalCN00,Agrawal2001,Gupta99}.

Finally, it would also be interesting to study how algorithms for mining functional dependencies~\cite{LPL00} 
or inclusion dependencies~\cite{DLP02} in databases might be exploited in our context. 
Many join operations (natural joins) are indeed carried out following  inclusion dependencies (concept of
foreign key). Discovering hidden dependencies within the data could thus help
us generating relevant indexes or materialized views without needing an input workload.

\bibliographystyle{latex8}
\bibliography{ideas03_darmont_j}

\begin{thebibliography}{10}\setlength{\itemsep}{-1ex}\small

\bibitem{agrawal93mining}
R.~Agrawal, T.~Imielinski, and A.~N. Swami.
\newblock Mining association rules between sets of items in large databases.
\newblock {\em SIGMOD Record}, 22(2):207--216, 1993.

\bibitem{agrawal94fast}
R.~Agrawal and R.~Srikant.
\newblock Fast algorithms for mining association rules.
\newblock In {\em 20th International Conference on Very Large Data Bases (VLDB
  1994), Santiago, Chile}, pages 487--499, 1994.

\bibitem{AgrawalCN00}
S.~Agrawal, S.~Chaudhuri, and V.~R. Narasayya.
\newblock Automated selection of materialized views and indexes in {SQL}
  databases.
\newblock In {\em 26th International Conference on Very Large Data Bases (VLDB
  2000), Cairo, Egypt}, pages 496--505, 2000.

\bibitem{Agrawal2001}
S.~Agrawal, S.~Chaudhuri, and V.~R. Narasayya.
\newblock Materialized view and index selection tool for {M}icrosoft {SQL}
  {S}erver 2000.
\newblock {\em 2001 ACM SIGMOD International Conference on Management of Data,
  Santa Barbara, USA}, 2001.

\bibitem{accidento}
K.~Aouiche.
\newblock Accidentology datamart schema and workload.
\newblock http://bdd.univ-lyon2.fr/download/charge-accidentologie.pdf, 2002.

\bibitem{Chaudhuri98}
S.~Chaudhuri.
\newblock Data mining and database systems: Where is the intersection?
\newblock {\em Data Engineering Bulletin}, 21(1):4--8, 1998.

\bibitem{chaudhuri97efficient}
S.~Chaudhuri and V.~R. Narasayya.
\newblock An efficient cost-driven index selection tool for {M}icrosoft {SQL}
  {S}erver.
\newblock In {\em 23rd International Conference on Very Large Data Bases (VLDB
  1997), Athens, Greece}, pages 146--155, 1997.

\bibitem{ChaudhuriN98}
S.~Chaudhuri and V.~R. Narasayya.
\newblock Autoadmin 'what-if' index analysis utility.
\newblock In {\em 1998 ACM SIGMOD International Conference on Management of
  Data, Seattle, USA}, pages 367--378, 1998.

\bibitem{DLP02}
F.~{De Marchi}, S.~Lopes, and J.-M. Petit.
\newblock Efficient algorithms for mining inclusion dependencies.
\newblock In {\em 8th International Conference on Extending Database Technology
  (EDBT 2002), Prague, Czech Republic}, volume 2287 of {\em LNCS}, pages
  464--476, 2002.

\bibitem{FinkelsteinST88}
S.~J. Finkelstein, M.~Schkolnick, and P.~Tiberio.
\newblock Physical database design for relational databases.
\newblock {\em TODS}, 13(1):91--128, 1988.

\bibitem{FrankON92}
M.~R. Frank, E.~Omiecinski, and S.~B. Navathe.
\newblock Adaptive and automated index selection in {RDBMS}.
\newblock In {\em 3rd International Conference on Extending Database Technology
  (EDBT 1992), Vienna, Austria}, volume 580 of {\em LNCS}, pages 277--292,
  1992.

\bibitem{Gupta99}
H.~Gupta.
\newblock {\em Selection and maintenance of views in a data warehouse}.
\newblock PhD thesis, Stanford University, 1999.

\bibitem{LPL00}
S.~Lopes, J.-M. Petit, and L.~Lakhal.
\newblock Efficient discovery of functional dependencies and {A}rmstrong
  relations.
\newblock In {\em 7th International Conference on Extending Database Technology
  (EDBT 2000), Konstanz, Germany}, volume 1777 of {\em LNCS}, pages 350--364,
  2000.

\bibitem{logexplorer}
{Lumigent Technologies}.
\newblock Log {E}xplorer for {SQL S}erver.
\newblock http://www.lumigent.com/products/le\_sql/le\_sql.htm, 2002.

\bibitem{ONeil95}
P.~O'Neil and G.~Graefe.
\newblock Multi-table joins through bitmapped join indices.
\newblock {\em SIGMOD Record}, 24(3):8--11, 1995.

\bibitem{oneil97improved}
P.~O'Neil and D.~Quass.
\newblock Improved query performance with variant indexes.
\newblock {\em SIGMOD Record}, 26(2):38--49, 1997.

\bibitem{pasquier99discovering}
N.~Pasquier, Y.~Bastide, R.~Taouil, and L.~Lakhal.
\newblock Discovering frequent closed itemsets for association rules.
\newblock In {\em 7th International Conference on Database Theory (ICDT 1999),
  Jerusalem, Israel}, volume 1540 of {\em LNCS}, pages 398--416, 1999.

\bibitem{close99is}
N.~Pasquier, Y.~Bastide, R.~Taouil, and L.~Lakhal.
\newblock Efficient mining of association rules using closed itemset lattices.
\newblock {\em Information Systems}, 24(1):25--46, 1999.

\bibitem{poess02}
M.~Poess, B.~Smith, L.~Kollar, and P.-A. Larson.
\newblock {TPC-DS}: Taking decision support benchmarking to the next level.
\newblock In {\em 2002 ACM SIGMOD International Conference on Management of
  Data, Madison, USA}, 2002.

\bibitem{tpcr99}
Transaction Processing Council.
\newblock {\em {TPC} Benchmark {R} Standard Specification}, 1999.

\bibitem{GV99}
S.~Vanichayobon and L.~Gruenwald.
\newblock Indexing techniques for data warehouses' queries.
\newblock Technical report, University of Oklahoma, School of Computer Science,
  1999.

\bibitem{WEI02}
G.~Weikum, A.~Monkeberg, C.~Hasse, and P.~Zabback.
\newblock Self-tuning database technology and information services: from
  wishful thinking to viable engineering.
\newblock In {\em 28th International Conference on Very Large Data Bases (VLDB
  2002), Hong Kong, China}, 2002.

\end{thebibliography}

\end{document}